\documentclass[twocolumn]{aastex62}
\usepackage{natbib}
\usepackage{graphicx}
\usepackage{txfonts}
\usepackage{comment}
\usepackage{color}
\newcommand{\chn}{\textit{Chandra}}

\newcommand{\ha}{H$\alpha$+[\ion{N}{2}]}

\received{Month day, 2018}
\revised{Month day, 2018}
\accepted{Month day, 2018}
\submitjournal{ApJ}

\shorttitle{Stormy weather in 3C 196.1}
\shortauthors{Ricci et al.}

\begin{document}

\title{Stormy weather in 3C 196.1: nuclear outbursts and merger events shape the environment of the hybrid radio galaxy 3C 196.1}

\correspondingauthor{Federica Ricci}
\email{fricci@astro.puc.cl}

\author[0000-0001-5742-5980]{F. Ricci}
\affiliation{Smithsonian Astrophysical Observatory, 60 Garden Street, Cambridge, MA 02138, USA}
\affiliation{Instituto de Astrof\'isica and Centro de Astroingenier\'ia, Facultad de F\'isica, Pontificia Universidad Cat\'olica de Chile, Casilla 306, Santiago 22, Chile}
\affiliation{Dipartimento di Matematica e Fisica, Universit\`a Roma Tre, via della Vasca Navale 84, 00146 Roma, Italy}
\author[0000-0002-3754-2415]{L. Lovisari}
\affiliation{Smithsonian Astrophysical Observatory, 60 Garden Street, Cambridge, MA 02138, USA}
\author[0000-0002-0765-0511]{R. P. Kraft}
\affiliation{Smithsonian Astrophysical Observatory, 60 Garden Street, Cambridge, MA 02138, USA}
\author[0000-0002-1704-9850]{F. Massaro}
\affiliation{Dipartimento di Fisica, Universit\`a degli Studi di Torino, via Pietro Giuria 1, I-10125 Torino, Italy}
\affiliation{Consorzio Interuniversitario per la Fisica Spaziale (CIFS), via Pietro Giuria 1, I-10125, Torino, Italy}
\author[0000-0002-5646-2410]{A. Paggi}
\affiliation{Dipartimento di Fisica, Universit\`a degli Studi di Torino, via Pietro Giuria 1, I-10125 Torino, Italy}
\author[0000-0003-0995-5201]{E. Liuzzo}
\affiliation{Istituto di Radioastronomia, INAF, via Gobetti 101, 40129, Bologna, Italy}
\author[0000-0002-5445-5401]{G. Tremblay}
\author[0000-0002-9478-1682]{W. R. Forman}
\affiliation{Smithsonian Astrophysical Observatory, 60 Garden Street, Cambridge, MA 02138, USA}
\author[0000-0002-4735-8224]{S. Baum}
\affiliation{University of Manitoba, Dept. of Physics and Astronomy, Winnipeg, MB R3T 2N2, Canada}
\affiliation{Center for Imaging Science, Rochester Institute of Technology, 84 Lomb Memorial Dr., Rochester, NY 14623, USA}
\author[0000-0001-6421-054X]{C. O'Dea}
\affiliation{University of Manitoba, Dept. of Physics and Astronomy, Winnipeg, MB R3T 2N2, Canada}
\affiliation{School of Physics \& Astronomy, Rochester Institute of Technology,
	84 Lomb Memorial Dr., Rochester, NY 14623, USA. }
\author[0000-0003-1809-2364]{B. Wilkes}
\affiliation{Smithsonian Astrophysical Observatory, 60 Garden Street, Cambridge, MA 02138, USA}




\begin{abstract}
We present a multi-wavelength analysis based on archival radio, optical and X-ray data
of the complex radio source 3C~196.1, whose host is the brightest cluster galaxy of a $z=0.198$ cluster.
HST data show \ha\ emission aligned 
	with the jet 8.4~GHz radio emission. 
	An \ha\ filament coincides with the brightest X-ray emission, the northern hotspot.
Analysis of the X-ray and radio images
reveals cavities located at galactic- and cluster- scales.
The galactic-scale cavity is almost devoid of 8.4~GHz radio emission 
and the south-western \ha\ emission is bounded (in projection) by this cavity. The outer cavity is 
co-spatial with the peak of 147~MHz radio emission, and hence we interpret this depression in X-ray surface brightness as being caused by a buoyantly rising bubble 
originating from an AGN outburst $\sim$280~Myrs ago.
A \chn\ snapshot observation allowed us to constrain the physical parameters of the 
cluster, which has a cool core with a low central 
temperature $\sim$2.8~keV, low central entropy index $\sim$13~keV~cm$^2$ and
a short cooling time of $\sim$500~Myr, which is $<0.05$ of the age of the Universe at this redshift.
By fitting jumps in
	the X-ray density
	we found 
	Mach numbers between 1.4 and 1.6, consistent with a shock origin.
We also found compelling evidence of a past merger, indicated 
by a morphology reminiscent of gas sloshing in the X-ray residual image. 
Finally, we computed the pressures, enthalpies $E_{cav}$ and jet powers $P_{jet}$ associated with the cavities: $E_{cav}\sim7\times10^{58}$~erg, $P_{jet}\sim1.9\times10^{44}$~erg~s$^{-1}$ for the inner cavity and $E_{cav}\sim3\times10^{60}$~erg, $P_{jet}\sim3.4\times10^{44}$~erg~s$^{-1}$
for the outer cavity.
\end{abstract}

\section{Introduction}

\label{s:intro}
Nuclear outflows from active galactic nuclei (AGN) have a dramatic impact on cosmic structure formation and evolution. 
These energetic outflows are invoked to
explain the anti-hierarchical quenching of star formation in
massive galaxies, the exponential cut-off at the bright end of the galaxy luminosity function, the 
black hole-host scaling relations and the quenching of cooling-flows in cluster cores 
\citep[e.g.][]{scannapieco05, begelman05, main17}.
The mutual radio galaxy/cluster interaction is significant for both: the intra-cluster medium (ICM) can change the jet propagation, while 
the great mechanical power of radio AGN can 
quench cooling in cluster cool cores.

One of the most important \chn\ results in
cluster science is the discovery of X-ray cavities and shocks 
\citep[e.g. see reviews from][]{mcnamaranulsen07,gitti12},
which are the smoking guns of the so-called radio- (or kinetic-, maintenance-) mode feedback \citep[\citealt{birzan08, cavagnolo10} for a review see][]{fabian12} of 
AGN, which are fed by the accretion of circumnuclear gas \citep[e.g.][]{gaspari15}.
The AGN energy released by the accretion process heats up or sweeps out the surrounding medium, 
reducing the accretion rate onto the black hole until it is switched off. 
When the gas cools down or it
is replenished, e.g. through a merger with another galaxy, accretion can start again. 
The new episode of accretion can eventually trigger the radio jets, closing the feedback loop.

The radio-mode feedback has been observed in 
cool core clusters \citep[MS0735.6+7421 \citealt{mcnamara09}; A2052 \citealt{blanton11}; A2597 \citealt{tremblay12}; M87][]{forman17}, in groups \citep[e.g. NGC 5813][]{randall15} and in isolated ellipticals \citep[e.g. NGC 4636][]{jones02} where the ICM is heated up by 
AGN jet-inflated plasma bubbles that buoyantly rise in the hot, X-ray emitting atmosphere \citep[][see also \citealt{boehringer93,churazov00,churazov01} for the first \textit{ROSAT} measurements in Perseus and Virgo]{churazov13, su17}.
As the bubbles rise, their energy is converted into motions of the X-ray emitting plasma which, eventually, is converted into thermal energy.

Our group has conducted over the last 8 years 
a \chn\ snapshot survey \citep[see][]{massaro10, massaro12,massaro13, massaro15, massaro18,stuardi18}
aimed at characterizing the X-ray radiation from 
jets, hotspots, nuclei and cluster emission of 
the most powerful, and probably most studied, 
radio galaxies known, the Third Cambridge Revised catalog \citep[3CR,][]{bennett62, spinrad85}.

Our \chn\ snapshot program allowed us to select 
targets for detailed 
analysis. 
3C 196.1 is very promising: it is
a hybrid morphology 
(both FR~I and FR~II, \citealt{fr74}) 
radio galaxy
embedded in a $kT\sim4$ keV galaxy cluster and has convincing morphological signatures
of radio galaxy/cluster interaction \citep{massaro12}, with structures in
the X-ray clearly related to ongoing and past outbursts of the radio galaxy.

3C 196.1 lies at $z=0.198$ 
within a cluster of galaxies 
CIZA J0815.4-0308
observed by
\textit{ROSAT} \citep{kocevski07}.
3C 196.1 has been classified in the optical as a low-excitation radio galaxy \citep[LERG,][]{buttiglione10}, and 
is associated with the brightest cluster galaxy (BCG), an elliptical cD galaxy, 
the dominant galaxy of a group of 14 others that lie within about 350 kpc from its core 
\citep{baum88}. 
\citet{madrid06} show that its near-infrared image is elliptical, presenting an 
elongated structure NE to SW, in the same direction as the inner scale jet. 
The same morphology is seen in the optical 
\citep{baum88,dekoff96}.
The optical color gradients of 3C 196.1 show periodic shells that
could reflect merging and galactic cannibalism \citep{zirbel96}.
The host galaxy has optical isophotes whose centroids clearly shift to the SW at small radii \citep{dekoff96}. 
This suggests that a merger has occurred several dynamical times ago.
In fact, stellar tidal debris from a merger wash out 
over a few dynamical times, while
shells in stellar surface brightness, which can manifest as isophote centroid offsets, can be long-lived remnants. 

In this work we investigate the physical conditions of the
3C 196.1 environment in order to shed light on 
ongoing interaction between the AGN and surrounding medium.
We present a multi-wavelength study based on radio, optical
and X-ray archival data of 3C 196.1, 
aimed at investigating the 
morphology and the nature of the multiphase emitting gas, 
unveil the impact of past AGN outbursts and dissect merger signatures.
The paper is organized as follows: in Sect. \ref{s:data} 
radio, optical and X-ray 
archival data are presented; 
Sect. \ref{s:analysis} describes the X-ray 
analysis, with imaging presented in Sect. \ref{sss:imaging}, 
spectral analysis carried out in Sect. \ref{sss:spec} and 
surface brightness profiles extracted and fitted in 
Sect. \ref{sss:sb}; 
in Sect. \ref{sss:edges} we adopted two edge-enhancement methods 
to localize small-amplitudes wriggles in the surface brightness;
in Sect. \ref{s:res} the results are presented and finally Sect. \ref{s:concl} is devoted to summary and conclusions. 
We assumed a flat cosmology with $H_0$ = 72~km~s$^{-1}$~Mpc$^{-1}$,
$\Omega_M$ = 0.27, and $\Omega_\Lambda$ = 0.73 \citep{dunkley09}. 
This means that for 3C 196.1, the angular scale is 3.197~kpc~arcsec$^{-1}$ and the luminosity distance is 946~Mpc.
For all the figures presented in this work, the standard astronomical orientation is adopted, i.e. north up and east to the left.

\section{Data}\label{s:data}

\begin{figure*}[t]
	\centering	
	
	\includegraphics[width=\textwidth]{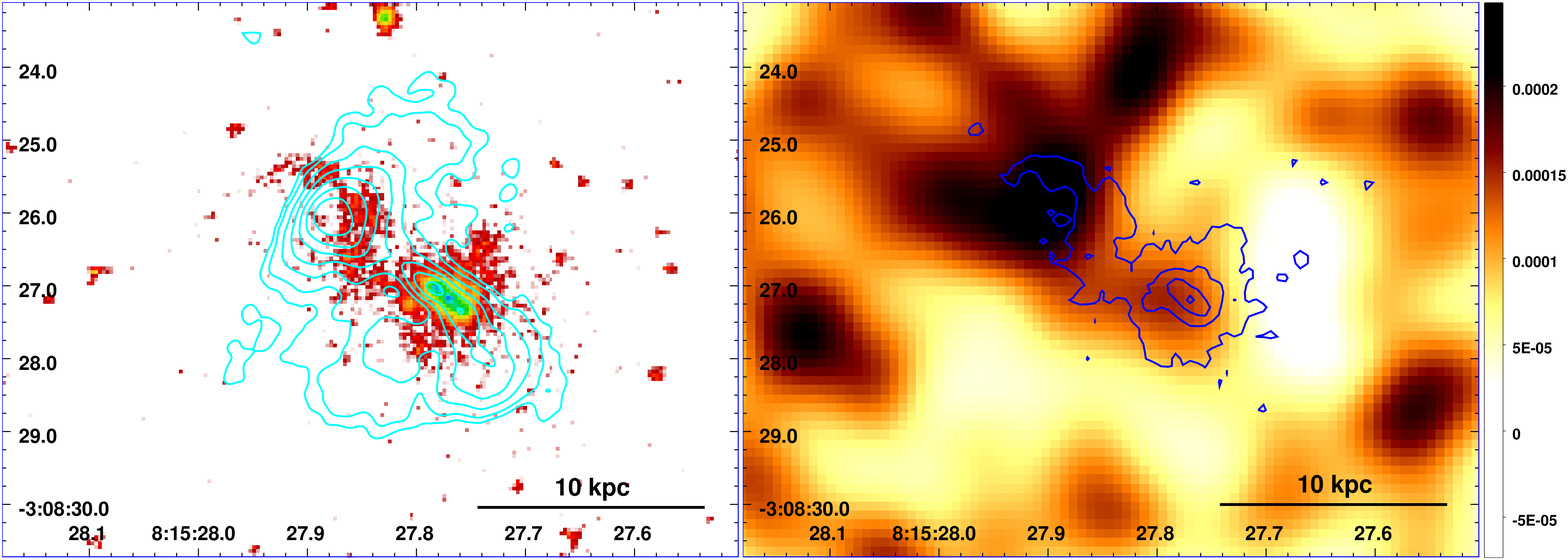}\vspace{0.05in} 
	
	\hspace{-0.26in}{\includegraphics[width=0.95\textwidth]{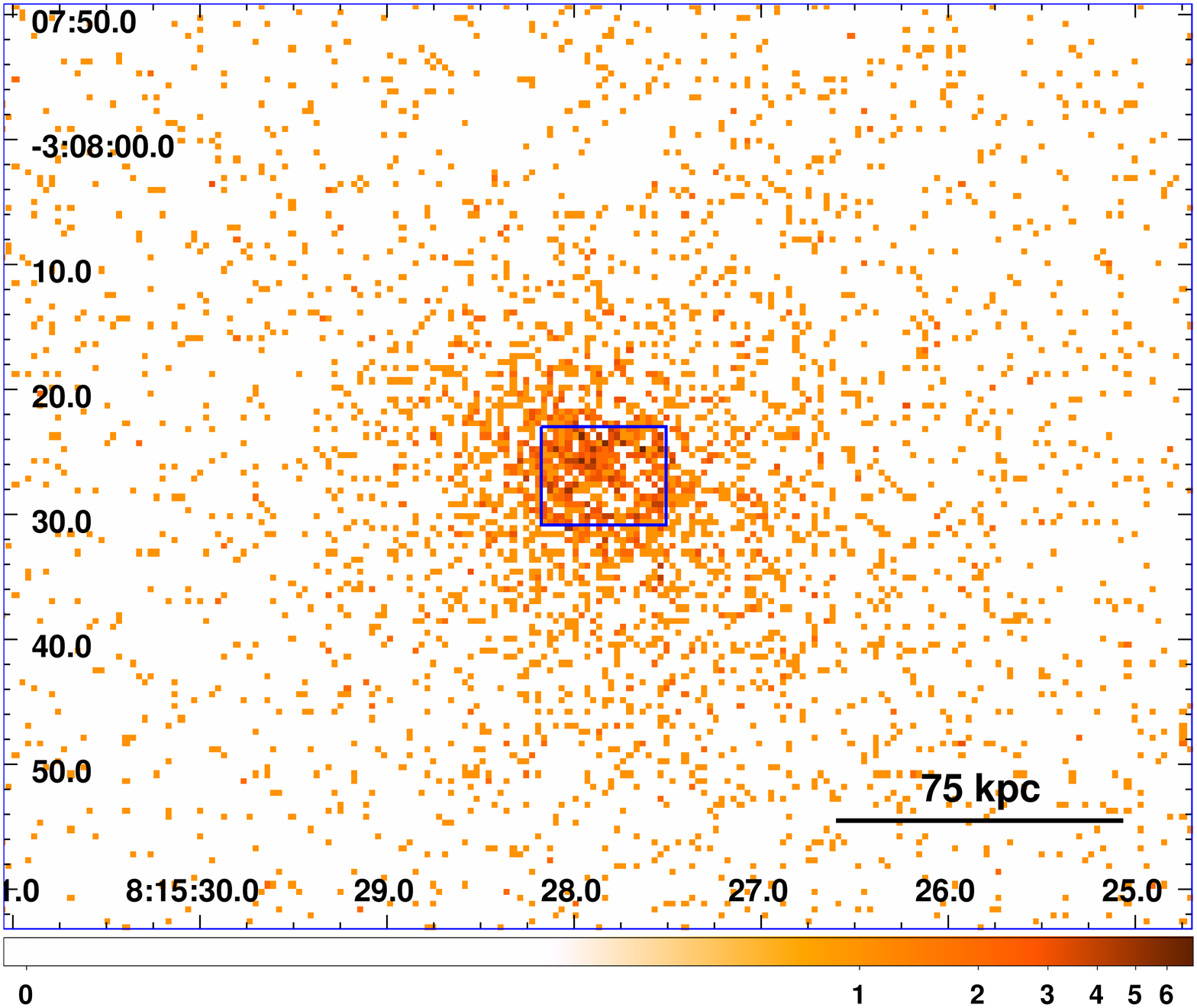}}
	\vspace{-0.05in}
	\caption{Images of the central BCG 3C 196.1 and of the surrounding cluster and ICM.
			Top panels show the
		 nuclear region of 3C 196.1  
		and share the same scale, with a FoV of 
		$\sim 10\times8$~arcsec$^2$ (i.e. $\sim 30\times25$~kpc$^2$).
		\textit{Top Left:} HST/ACS observation of 3C 196.1 from \citet{tremblay09}, with 8.4 GHz VLA contours superimposed. 
		The HST data have been continuum subtracted, to show only the \ha\ emission.
		The 8.4~GHz contours 
		start at $\sim 3 \sigma$, i.e. 0.275~mJy, and increase with a square-root scale; the clean beam is 
		(0$\farcs$3$\times$0$\farcs$3, 0$^\circ$).
		The \ha\ emission has been aligned with the 8.4~GHz VLA data (shift $\lesssim$1$\farcs$2, see text for details). 
		An \ha\ filament extends across the NE radio lobe.  
		The \ha\ is aligned with the radio emission \citep[so-called alignment effect, e.g. ][]{baum88}.
		\textit{Top Right:} The \chn\ 0.7-2~keV emission is shown with the \ha\ contours superimposed. No shift was applied to
		align the images. The \chn\ data have 0$\farcs$123 pixel scale and have been smoothed with a Gaussian kernel of 1$\arcsec$. Depressions in surface brightness are in white. The \ha\ contours start at 
		$0.378\times 10^{-19}$~erg~s$^{-1}$~cm$^{-2}$~\AA$^{-1}$
		and increase with a square-root scale. The SW \ha\ emission is bounded to the W, S and SE by regions of low X-ray surface brightness.
		\textit{Bottom:} The 0.5-10~keV \chn\ image, showing an FoV of $\sim 93\times 73$~arcsec$^2$ (i.e. $\sim 297\times234$~kpc$^2$), with native pixel size. 
			The blue box at the center marks the region of 3C196.1 shown in the top panels.
			The logarithmic color scale shows the counts per pixel.}
	\label{fig:halpha}
\end{figure*}

\begin{figure*}
	\centering	
	
	\includegraphics[width=\textwidth]{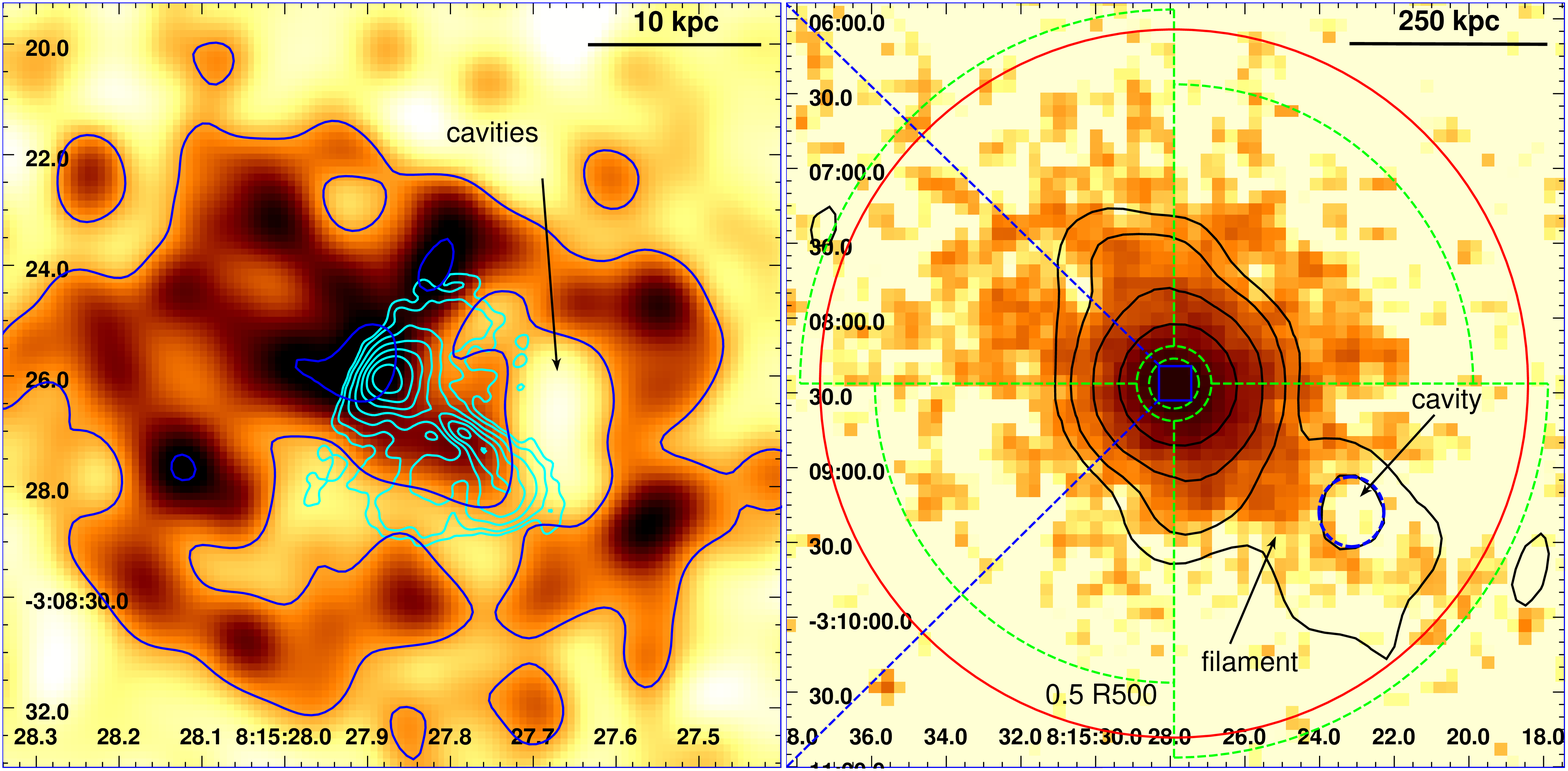}
	\caption{\chn\ 0.7-2~keV energy filtered and point-sources subtracted images of the central BCG 3C 196.1 and of the surrounding cluster and ICM. Regions of low surface brightness are in white.
	\textit{Left:} Inner $\sim 13\times 14$~arcsec$^2$ (i.e. $\sim 44\times45$~kpc$^2$) of the BCG, centered on the brightest X-ray peak. The image is shown with a 0$\farcs$123 pixel size and has been smoothed with a 1$\arcsec$ Gaussian. VLA 8.4~GHz contours are overlaid in cyan (for details see Fig. \ref{fig:halpha}), while the X-ray contours are shown in blue.
	The lowest blue contour encloses a region of low surface brightness, a ``butterfly-shaped'' cavity (marked with a black arrow).
	The brightest X-ray peak is co-spatial with the northern radio lobe and could be a hotspot.		
	\textit{Right:} The BCG and the cluster are shown up to the outskirts, with a FoV of 
	$\sim$5$\times$5~arcmin$^2$ (i.e. $\sim$960$\times$960~kpc$^2$). 
	Pixel size is 5$\arcsec$, Gaussian smoothing with FWHM=5$\arcsec$. 
	The central blue box is $\sim 13\times 14$~arcsec$^2$ and contains the zoom in shown in the left panel.
	GMRT 147~MHz contours are overlaid in black starting at $\sim$3$\sigma$, i.e. 39~mJy with a logarithmic scale. The clean beam is  (26$\farcs$96$\times$25$\arcsec$, 0$^\circ$).
	The arrow marks a large filament which has a projected extension of $\sim$2$'$ (i.e. $\sim$~384~kpc). 
	It may also be the bottom of a large cavity (highlighted by the dashed blue ellipse), which is 
	co-spatial with the peak of the SW low-frequency diffuse emission.
	The red circle shows the estimated 0.5$R_{500}$.
	Extraction regions used in Sect. \ref{sss:spec} are shown with green dashed lines.}
	\label{fig:xray}
\end{figure*}

\subsection{Radio data}
We used 
high frequency 8.4~GHz Very Large Array (VLA) archival data to investigate the 
core region of the galaxy,
whereas we used low frequency 
147~MHz Giant Metrewave Radio Telescope (GMRT) 
archival data to study the diffuse 
emission in the cluster outskirts ($\sim$~500~kpc, i.e. $\sim$0.5~$R_{500}$\footnote{$R_{500}$ defines the radius at which the over-density 
is equal to 500	times the critical density of the Universe at the cluster redshift.}). 
These radio frequencies sample different spatial scales due to the angular resolution that each radio band can achieve.
The high frequency 8.4~GHz VLA data shown in Fig. \ref{fig:halpha} 
(left panel, contours on top of the optical image)
has been 
observed on 2 July 1995 with the VLA
in configuration A,
the restoring beam is (0$\farcs$3$\times$0$\farcs$3,0$^\circ$),
the rms is $9.13\times10^{-2}$~mJy/beam and the brightest peak is at $\sim 1.37\times10^{-2}$~Jy/beam. 
The 8.4~GHz flux of the NE (SW) radio lobe is 71~(32)~mJy.

The 147~MHz GMRT data has been 
retrieved from the 
TIFR GMRT Sky Survey Alternative Data Release 
\citep[TGSSADR][]{intema17} 
website\footnote{http://tgssadr.strw.leidenuniv.nl/doku.php}.
The TGSS ADR ID number is R26D31 and is a mosaic image of 5 deg square.
The 147~MHz emission contours are superposed on the X-ray emission in Fig. \ref{fig:xray} (right panel, black contours). 
The restoring beam is (26$\farcs$96~$\times$~25$\arcsec$, 0$^\circ$),
the rms is 0.13~mJy/beam and the brightest peak is at $\sim$~16.9~Jy/beam.
The radio flux of the SW diffuse emission 
is 0.264~Jy at 147~MHz.

\subsection{HST observation}
The optical archival data were observed on 3 December 2006 during Cycle 15 
with the Advanced Camera for Surveys (ACS) aboard HST, 
snapshot program 10882 (PI: Sparks), originally published by \citet[][we refer the reader to this work for details on data reduction and calibration]{tremblay09}. 
The narrowband 
ACS ramp filter F782N was used to
trace the restframe optical \ha\ emission line complex. 
The HST data have been continuum subtracted to isolate the \ha\ emission.
In Fig. \ref{fig:halpha} we report the central part of the radio galaxy 3C 196.1 imaged with HST/ACS with the high-frequency radio emission at 8.4 GHz superimposed (cyan contours). 
The HST image was registered in order to align the optical isophotal peak with the southern jet component. The shift of the HST image was $\lesssim$1$\farcs$2.
The overall \ha\ emission spans $\sim$10~kpc projected distance, inside the galaxy, 
starting from the core and extending along the radio jet.
The \ha\ emission is aligned with the 8.4 GHz radio emission.
Similar alignments of radio with optical emission lines have been observed in many radio galaxies and is known as 
``alignment effect'': in powerful radio galaxies,
the optical line emitting gas (T$\sim$10$^4$~K) in the
proximity of the nuclei has been found to be spatially aligned with the radio jet axes 
on kpc-scales \citep{fosbury86, hansen87, baum88, baum90,devries99, tremblay09}.
The alignment effect is generally explained as due to shocks induced by the jet propagation and by AGN photoionization \citep{baum89, dopita95, dopita96, best00}.
 
There is also a filament to the NE which drapes across the northern radio lobe. 
This filamentary structure is not seen in [\ion{O}{3}], which 
shows high-excitation emission in two localized hotspots co-spatial with the elongated bright core \citep[see Fig. 7 in][]{tremblay09}.

\subsection{\chn\ observation}\label{ss:datax}
3C 196.1 was observed during \chn\ AO12, 
OBSID 12729 on 11 February 2011, with a nominal 8~ks exposure. 
The ACIS-S back-illuminated chip was used in VERY FAINT MODE with standard frame times (3.2~s). 
The source was positioned on ACIS-S3. 
Data reduction has been performed following the standard reduction procedure described in the \chn\ Interactive Analysis of Observations 
\citep[CIAO, ][]{fruscione06}
threads\footnote{http://cxc.harvard.edu/ciao/guides/index.html}, using CIAO v4.9 and the \chn\ CalDB version 4.7.4. 
Level 2 event files were generated using the 
$\tt{chandra\_repro}$ task. Events were filtered for grades 0, 2, 3, 4, and 6. The light curve was extracted and checked for high background intervals, the actual live time is 7.93~ks.	
The resulting 0.5-10~keV \chn\ image in photon counts with native pixel size is shown in the bottom panel of Fig. \ref{fig:halpha}, in order to show the structures in the original data without 
any Gaussian smoothing or image processing.

\section{X-ray Analysis}\label{s:analysis}
\subsection{Imaging}\label{sss:imaging}

Fig. \ref{fig:xray} presents the diffuse X-ray emission in the energy range 0.7-2~keV, with pixel size 0$\farcs$123 (left) and 5$\arcsec$ (right).
Point sources were detected with the CIAO task $\tt{wavdetect}$
(with 1, 2, 4, 8, 16 and 32 pixels sequence of wavelet scales), adopting a probability of spurious detection of $10^{-6}$, and then replaced with local background.
Five point sources were detected, all located outside the regions considered in our subsequent analysis. Thus, our results are not affected by the removal of point sources.
The images have been smoothed with a Gaussian kernel of 1$\arcsec$ (left) and 5$\arcsec$ (right). 
The left panel of Fig. \ref{fig:xray} shows the central $\sim$~44$\times$45~kpc$^2$ region together with 8.4~GHz contours (cyan) and X-ray contours (blue).
The radio source is oriented along the steepest gradient in the X-ray emission.
The X-ray surface brightness is complex and is not symmetric, extended along a position angle PA\footnote{Positive direction is counterclock-wise 
	north through east.}$\sim$40$^\circ$ that matches the 
optical major axis of the BCG stellar isophotes \citep[see Fig.~32 in][]{dekoff96}.
The image
shows the presence of sharp surface brightness edges in the
central regions of the cluster.
The lowest blue contour marks a system of ``butterfly-shaped'' surface brightness deficits (see black arrow in Fig. \ref{fig:xray}), 
also reported by \citet{massaro12} as a ``ghost cavity''. 
The southern jet of the 8.4~GHz source (cyan contours) is bounded (in projection) to the W and E by these
``butterfly-shaped'' deficits of X-ray radiation. 
This cavity is mostly devoid of high frequency radio emission.
The best description of this source with the available 0$\farcs$3 resolution is an \textit{hymor} 
\citep[hybrid morphology radio source][]{gopal-krishna00}.
The SW side is jet-like (i.e., FR~I) whereas the NE side appears to be a 
classical FR~II lobe with a brightness enhancement toward the edge, 
the X-ray peak marking the location where the 
NE jet impacts the higher density ICM. 
Several radio sources have been observed to exhibit a mixed morphology 
\citep{kaiser07, kharb10,kapinska17}, suggesting the FR~I/II dichotomy is at least partly due to environmental effects.

On the central panel of Fig. \ref{fig:halpha} the central region is further magnified, with a field of view (FoV) of $\sim$ 30$\times$25~kpc$^2$. The \ha\ emission is 
overlaid with blue contours. No shift was applied to register the \chn\ and HST images.
The SW optical line emission appears to be confined by the 
butterfly-shaped X-ray cavity. The \ha\ NE filamentary emission is co-spatial
with the northern hotspot.

Also at larger scales, the diffuse X-ray emitting gas distribution is asymmetric, elongated north-east to south-west, as shown in the
right panel of Fig. \ref{fig:xray}, where the system 
is imaged 
up to $\sim 0.5 R_{500}$, with a FoV of $\sim$1$\times$1~Mpc$^2$.
The image shows that the peak of the X-ray emission is centred on the 
	BCG, and that there is a gradient in the surface brightness distribution, 
	with the lowest surface brightness region extending up to few hundreds of kpc.
Right panel of Fig. \ref{fig:xray} presents low frequency 147~MHz GMRT contours in black.
The peak of the low-frequency 147~MHz emission (in the SW extension) is co-spatial with a region of low X-ray surface brightness (see dashed blue ellipse), that could be a large cavity filled with a low-frequency radio bubble.
In the outer region there appears an X-ray filament (see black arrow) which has a projected extension of $\sim$2$\arcmin$, corresponding to $\sim$~384~kpc. It may also outline the southern border of the aforementioned large cavity.

To further explore the underlying physical state of the gas, we 
derive the temperature in both the nuclear region and the cluster
outskirts in Sect. \ref{sss:spec}, and we analyse the surface brightness profiles 
in Sect. \ref{sss:sb}.

\subsection{Spectral analysis}\label{sss:spec}
We performed a temperature analysis of the 
cluster, from the inner core 
up to $\sim0.5R_{500}$.
First we analysed the central core, 
considering two circles of radius $r=10\arcsec$ and $15\arcsec$,
both
centred on the centroid of the X-ray emission (RA$_{J2000}$=8:15:27.900 and DEC$_{J2000}$=-3:08:26.265). 
These two central regions are called \textit{core 0} and \textit{core 1} in Table \ref{tab:xfit}.
We then analysed the cluster by dividing the emission into four sectors evenly spanning the 360 degrees.
The radial extents of these four regions are 15$\arcsec$-2$\arcmin$ for the NW and SE sectors (since in the SE direction the 
S3 chip ends at that radius) and 15$\arcsec$-2$\farcm$5 in the NE and SW sectors
(see the 6 green dashed regions in the right panel of Fig. \ref{fig:xray}). 
We also considered the whole region within  $r=2\arcmin$ to derive the global temperature.

We extracted the spectra in these regions 
using the CIAO task $\tt{specextract}$, thereby automating the
creation of count-weighted response matrices. 
Background correction was taken into account using the proper blank sky field event file, 
which was created using the CIAO task 
$\tt{blanksky}$.
Background spectra were extracted from the blank sky observation that was re-projected to match
the observation.
The background-subtracted spectra were then filtered in energy between 0.5 and 7~keV, binned using a 20 count threshold  
and fit  
using iterative $\chi^2$ minimization techniques with $\tt{xspec}$ version 12.9.1 \citep{arnaud96}.

The model adopted is $\tt{phabs\times APEC}$ with
column density fixed to the Galactic value $N_H = 5.81 \times 10^{20}$~cm$^{-2}$. 
We fitted each region twice, first leaving the abundance free to vary and 
then fixing it to 0.3 times the solar value, which is often found in cluster atmospheres \citep[see e.g.][]{owers09}.
Abundances were computed taking as reference the solar metallicity as reported by \citet{asplund09}. 
Best-fit values are listed in Table \ref{tab:xfit}.

\begin{table}
	\caption{Results of the spectral fitting analysis of the 0.5-7~keV \chn\ data.
		The $\tt{XSPEC}$ model used is $\tt{phabs\times APEC}$, with $N_H$ column fixed to 
		Galactic value and $z$ fixed to 0.198.
		Prior to extraction, the data were filtered for flaring events, 
		and compact point sources were removed. 
		Spectra were re-binned using a 20-count threshold. 
		Symmetric 1$\sigma$ confidence intervals are shown on fit parameters.}\label{tab:xfit}              
	\begin{center}                                      
		\begin{tabular}{l c c c c l}          
			\hline\hline                        
			\smallskip 
			
			Region	&N		&$kT$			 &Z		&APEC norm				 & $\chi^2_\nu$($\nu$)		\\
			$[r_{in}$,$r_{out}]$		&				&[keV]			 &[Z$_\odot$]		&[10$^{-3}$cm$^{-5}$]						 & \\
			(1) & (2) & (3) & (4) & (5) & (6) \\
			\hline 
			core 0&1115			&2.88$\pm$0.19 &0.80$\pm$0.24 &1.090$\pm$0.090  & 0.99(44)\\
			$[0,10\arcsec]$&							&2.81$\pm$0.22 &0.3$^\dagger$  		&1.297$\pm$0.046	 & 1.13(45)\\
			&			&&&&\\
			core 1&1534			&3.18$\pm$0.17 &0.98$\pm$0.23 &1.406$\pm$0.097  & 0.93(57)\\
			$[0,15\arcsec]$&							&3.14$\pm$0.21 &0.3$^\dagger$   		&1.718$\pm$0.053		 & 1.16(58)\\
			&			&&&&\\
			NW 			&809				&5.23$\pm$0.99 &0.24$\pm$0.35 &0.791$\pm$0.075	 & 1.00(33)	\\
			$[15\arcsec ,2\arcmin]$&			&5.24$\pm$0.97 &0.3$^\dagger$   			&0.781$\pm$0.036	 & 0.97(34)\\
			
			&				&&&&\\
			NE			&959				&5.15$\pm$0.82 &1.12$\pm$0.65 &0.714$\pm$0.080 	&0.98(39) 	\\		
			$[15\arcsec, 2\farcm5]$&	&4.85$\pm$0.88 &0.3$^\dagger$  			&0.833$\pm$0.038	&1.02(40)\\
			
			&				&&&&\\
			SE 	&590				&4.66$\pm$0.92 &0.99$\pm$0.86 &0.399$\pm$0.063 	&1.20(25)\\
			$[15\arcsec,2\arcmin]$&	&4.37$\pm$0.94 &0.3$^\dagger$ 			&0.462$\pm$0.027	&1.20(26)\\
			
			&				&&&&\\
			SW 	&1194				&7.63$\pm$1.80 &0.16$\pm$0.35 &0.975$\pm$0.078 	&0.97(50)\\
			$[15\arcsec,2\farcm5]$&	&7.46$\pm$1.68 &0.3$^\dagger$ 			&0.956$\pm$0.035	&0.95(51)\\
			
			&				&&&&\\
			All	&4228				&4.24$\pm$0.24 &0.73$\pm$0.14 &4.01$\pm$0.14 	&1.10(154)\\
			$[0,2\arcmin]$&	&4.25$\pm$0.27 &0.3$^\dagger$  			&4.412$\pm$0.084	&1.16(155)\\

			\hline                                             
		\end{tabular}
	\end{center} 
	\normalsize	
	\tablecomments{Columns are: (1) extraction region, with range $[r_{in}$,$r_{out}]$, centered at RA$_{J2000}$=8:15:27.900 and DEC$_{J2000}$=-3:08:26.265 (see Fig. \ref{fig:xray}); (2) net-counts; (3) the best-fit temperature of the $\tt{APEC}$ model in keV units;
		(4) the abundance, in solar units; (5) the $\tt{APEC}$ normalization; 
		(6) the reduced chi square $\chi^2_\nu$, with the degrees of freedom $\nu$ in parentheses.\\
		$^\dagger$  Parameter has been frozen.}	
	
\end{table}

\begin{figure*}[h!]
	\centering

	\includegraphics[width=\columnwidth]{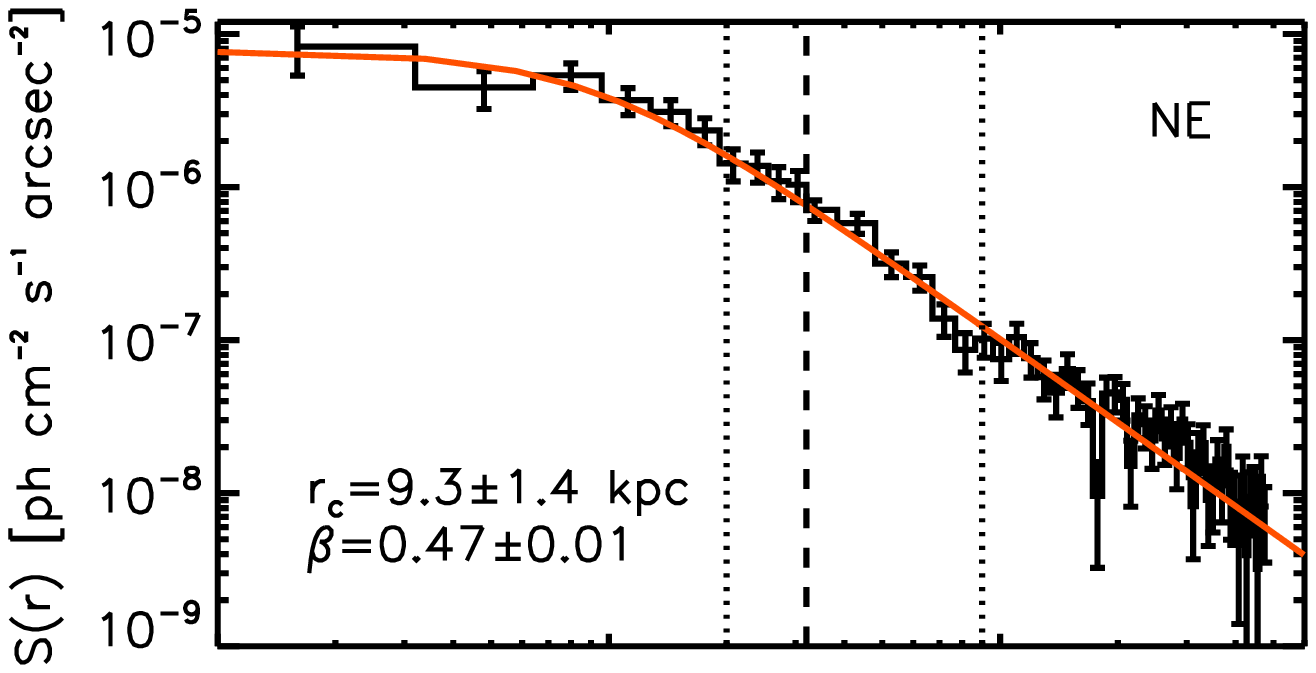}\hspace{-.8in}
	\includegraphics[width=\columnwidth]{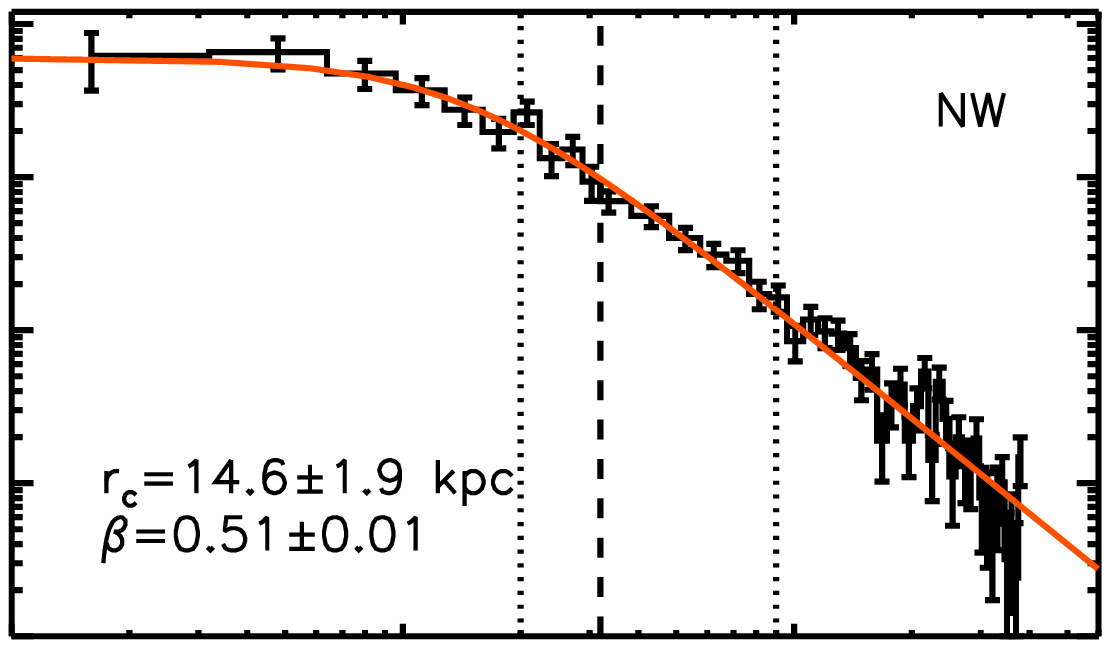}\vspace{-.67in}
	
	\includegraphics[width=\columnwidth]{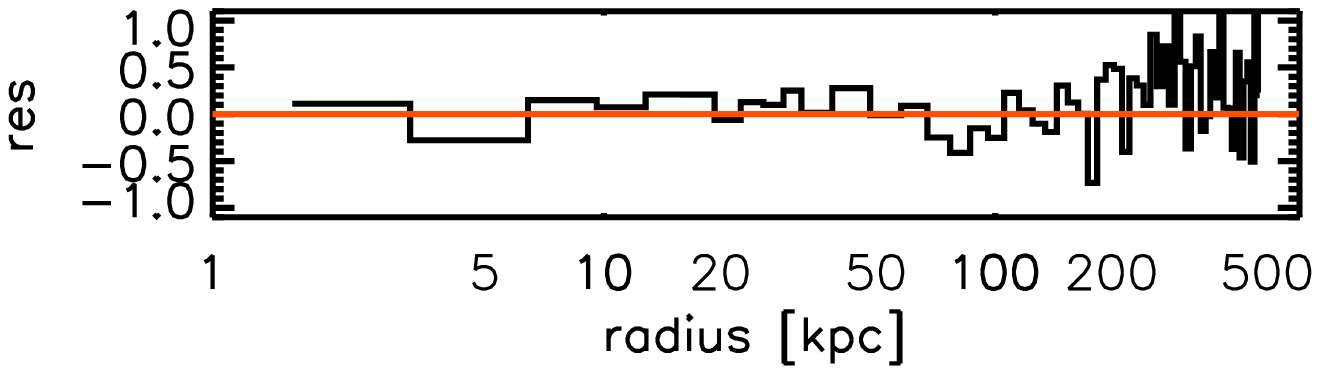}\hspace{-.8in}
	\includegraphics[width=\columnwidth]{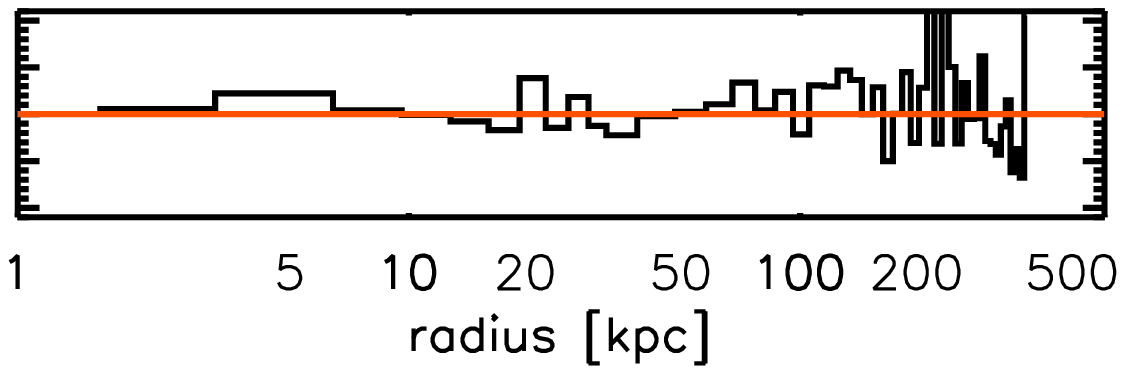}\vspace{-.25in}
	
	\includegraphics[width=\columnwidth]{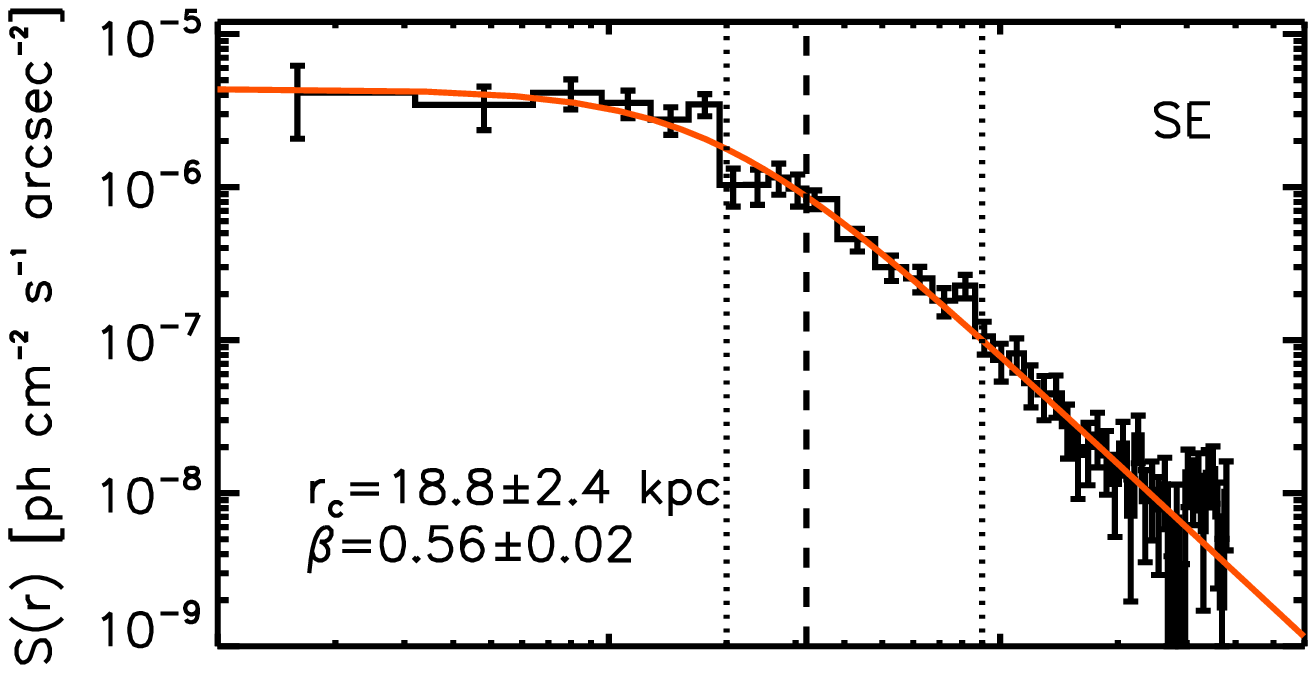}\hspace{-.8in}
	\includegraphics[width=\columnwidth]{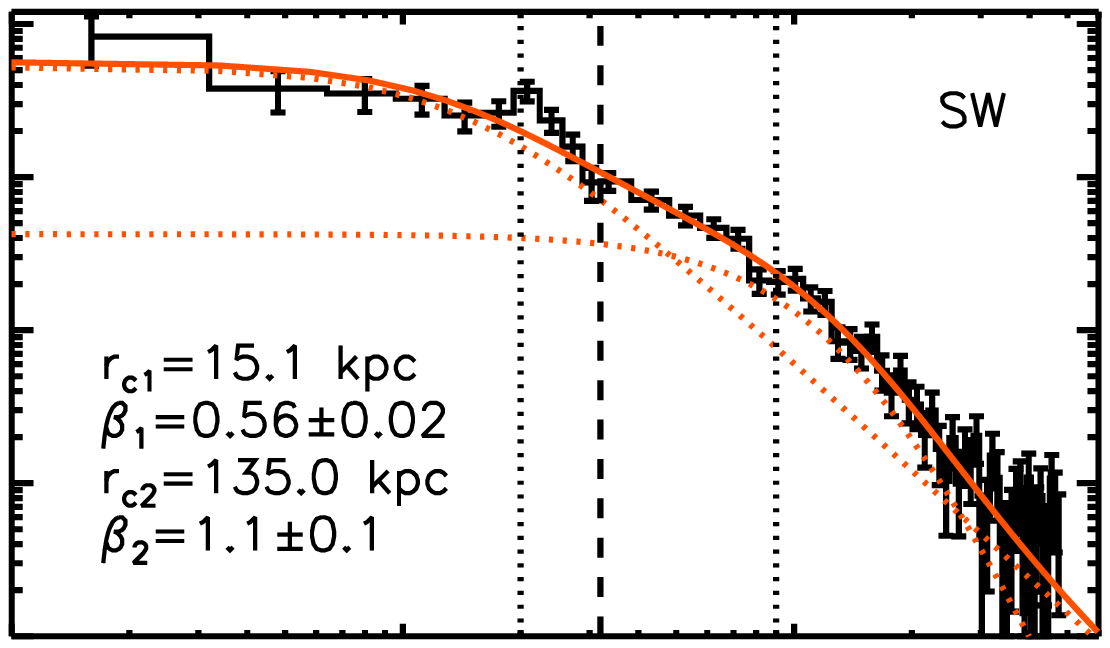}\vspace{-.67in}
	
	\includegraphics[width=\columnwidth]{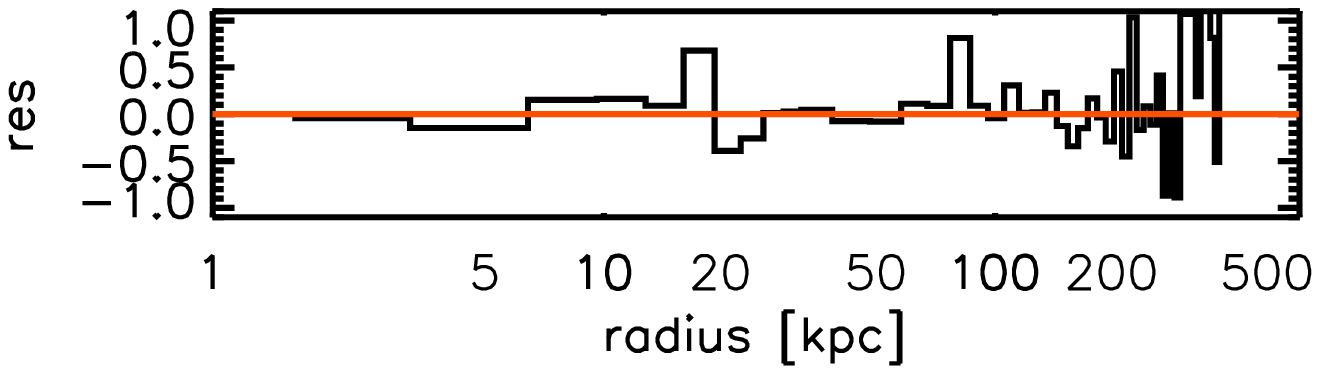}\hspace{-.8in}
	\includegraphics[width=\columnwidth]{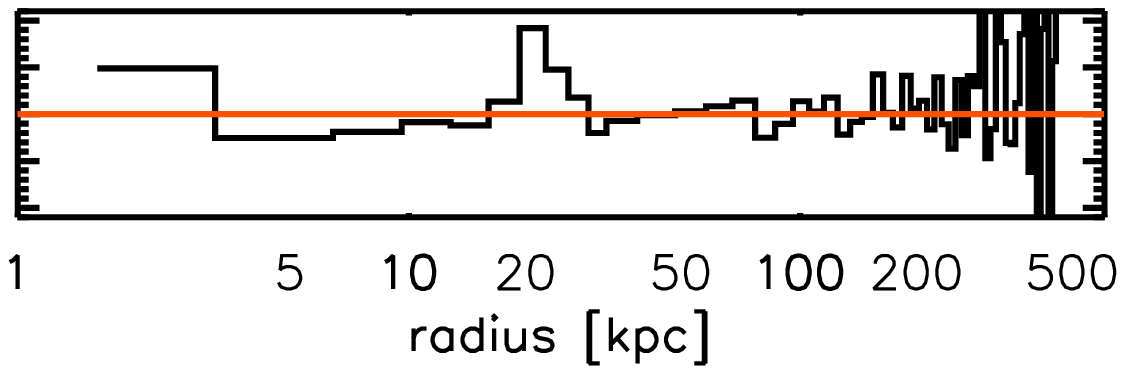}\vspace{-.25in}
	
	\includegraphics[width=\columnwidth]{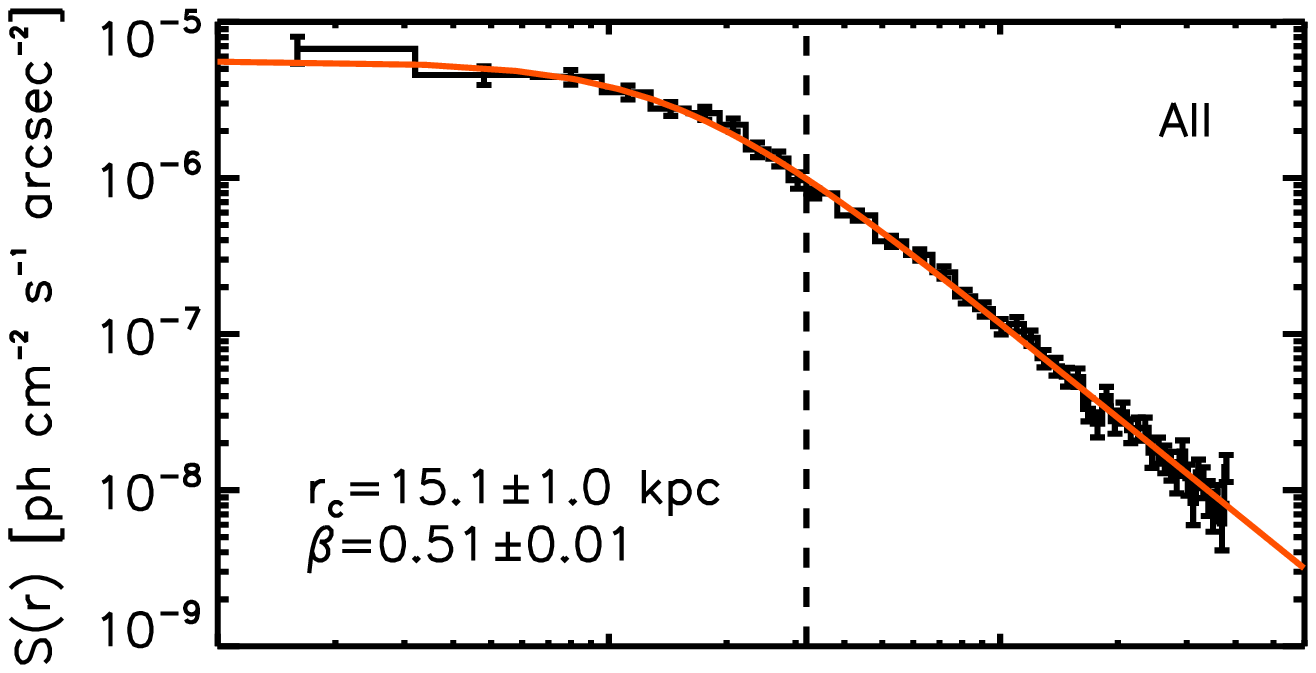}\hspace{0.78\columnwidth}\vspace{-.67in}
	
	\includegraphics[width=\columnwidth]{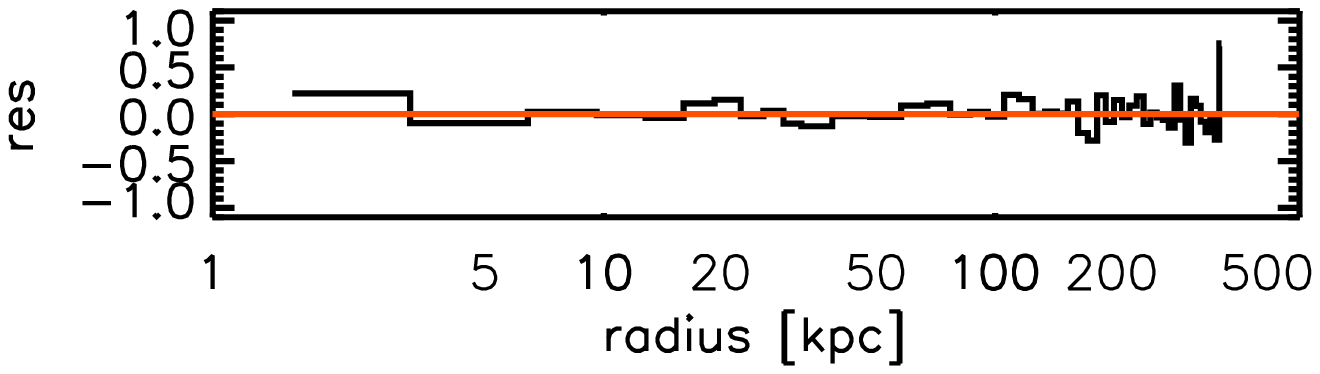}\hspace{0.78\columnwidth}

	\caption{Exposure-corrected and background-subtracted surface brightness profiles of four sectors, evenly spanning the 360 degree, and of an azimuthal region, extending up to 2$\arcmin$. 
		Bright point sources were excluded.   
		Radial bins are 3$\arcsec$ wide, with the exception of 
		the inner 10$\arcsec$, which were spanned by 1$\arcsec$-bins. 
		The radial profiles have been computed in 0.7-2~keV band, which is less sensitive to the temperature. 
		Parameters of the $\beta$-models are also reported. The dashed line marks the inner core region $r=10\arcsec$, which has a lower temperature of $\sim3$~keV (see region \textit{core 0} in Table \ref{tab:xfit}). 
		The dotted lines show the location of possible
		density jumps, reflected as change of slopes in the surface brightness profiles.
		Lower panels show the residuals of the fit.}
	\label{fig2}
\end{figure*}

\subsection{Surface brightness profiles}\label{sss:sb}
We extracted surface brightness profiles 
by considering an azimuthal area up to $r=2\arcmin$
and again in four sectors with outer radii 2$\arcmin$ (NW and SE) or 2$\farcm$5 (NE and SW).
All these five regions were spanned by 3$\arcsec$ wide radial bins, apart 
from the inner core region (up to 10$\arcsec$) which was divided into 1$\arcsec$-wide bins.  
In the surface brightness profile analysis, the \chn\ event file was 
filtered in energy between 0.7-2~keV band, which is less sensitive 
to the gas temperature.
It was then
exposure-corrected, background-subtracted and bright point sources were excluded.
The same point sources area was also removed from the blank sky field.
Monochromatic exposure maps were computed with nominal energies of
1.35~keV.
The background-subtracted and exposure-corrected radial profiles 
are shown in Fig. \ref{fig2}.

Given the limited counts in our short snapshot observation, 
we fit the surface brightness profiles at projected distance $r$
with a $\beta$-model \citep{cavaliere76}
\begin{equation}\label{eq:beta}
S(r) =  S_{0}  \left[ 1+ \left( \frac{r}{r_{c}} \right)^2 \right]^{-3\beta +0.5}\, ,
\end{equation}
where $S_{0}$ is the central X-ray surface brightness and $r_c$ is the core radius.
We adopted the $\tt{IDL}$ package $\tt{MPFIT}$ \citep{mpfit}, 
which uses the Levenberg-Marquardt technique to determine the least-squares best fit. 
Best-fit models are reported as red solid curves in Fig. \ref{fig2},
together with best-fit parameters.
Residuals are shown in the lower panels of Fig. \ref{fig2}.

The F-test was then carried out in order to quantitatively verify
whether a two $\beta$-model\footnote{A two $\beta$-model is defined as the linear combination of two different $\beta$-models as given in eq. \ref{eq:beta}.} was to be preferred,
adopting a significance threshold of 0.05.
The two $\beta$-model was always rejected, with the exception of the SW sector.

The ICM in galaxy clusters is typically well fitted by
models with $\beta \sim  0.6$ \citep[][]{sarazin86, ettori09}, 
which steepens with increasing core radius as the two parameters are 
strongly correlated \citep{morandi15}. 
Our results are in good agreement, as we find 
that the average surface brightness (azimuthally extracted up to 2$\arcmin$)
is well fitted by a single $\beta$-model with $\beta=0.51\pm0.01$.
The other three sectors give a rather consistent picture, with
$\beta$ ranging from 0.47$\pm$0.01 in the NE sector to 0.56$\pm$0.02
in the SE sector. 
To properly fit the SW profile, 
we fixed the core radius to $r_{c,1}\sim15$~kpc (as found from the 
average azimuthal fit) and $r_{c,2}\sim 135$~kpc, and the resulting 
best-fitting slopes are: $\beta_1 = 0.56\pm 0.02$, consistent with 
the fit results of the other four regions, and  
$\beta_2 = 1.1 \pm 0.1$.

The surface brightness profile, closely tracing the underlying
	3-d distribution of the gas density, should strictly follow 
an isothermal profile. Thus
a small deviation from the theoretical profile, visible as a change of slope in the 
surface brightness radial profile, reflects a discontinuity in the underlying gas density. 
The surface brightness profiles plotted in Fig. \ref{fig2}
indeed show these changes of slopes,
marked with dotted lines,
at $\sim$~20~kpc, 
which were already visible as sharp 
surface brightness
``edges'' in the central 44$\times$45~kpc$^2$ FoV in Fig. \ref{fig:xray}, left panel.
There is a discontinuity in the slope of the surface brightness radial profile
also in the outer region,
around $\sim$~90~kpc, that was not evident in Fig. \ref{fig:xray}. 

To constrain the discontinuity conditions,
in the most discrepant cases, e.g. NE and SE,	
we fit the surface brightness profile within each region
assuming spherical symmetry for the gas density and constant
gas temperature. The significance
of the density discontinuity was quantified by
modelling the surface brightness profile across the contact discontinuity with a
broken power law density model \citep[see Appendix of][]{owers09}. Given the limited counts of our snapshot observation, we fix the location of the density discontinuity to help the fitting minimization procedure.

For temperatures greater than 2~keV, the discontinuity in the slope of the surface brightness profile is 
related to the discontinuity in the gas density as 
\begin{equation}
\frac{n_{e,1}}{n_{e,2}} \approx \sqrt{\frac{S_{1}}{S_{2}}} \, ,
\end{equation}
where $n_{e,1}$, $n_{e,2}$ are the densities on either side of the edge,
and $S_{1}$, $S_{2}$ are the amplitudes of the corresponding surface brightness values. 
These profiles reveal density ratios of 
$1.61\pm0.51$ at a distance of 21~kpc, 
and $1.77\pm0.30$ at a distance of 90~kpc in the SE sector, and $1.80\pm0.27$ at distance of 85~kpc in the NE region.
The position of the outer jumps 
are
quite symmetric between the two sides, since
they are at 90 and 85~kpc SE and NE respectively.
Assuming the density discontinuities are associated with shocks, the Mach number $\mathcal{M}$ is given by the Rankine-Hugoniot jump condition: 
\begin{equation}\label{eq:mach}
\frac{n_{e,d}}{n_{e,u}} = \frac{\gamma + 1}{ \gamma -1 +2\mathcal{M}^{-2}}
\end{equation}
where $n_{e,d}$ and $n_{e,u}$ are the densities downstream and upstream of the shock, respectively.
Assuming an ideal gas with ratio of specific heats $\gamma=5/3$ and the density ratios measured above, 
the resulting Mach numbers are:
$\mathcal{M}=1.42\pm0.38$ (SE, r=21~kpc); 
$\mathcal{M}=1.54\pm0.24$ (SE, r=85~kpc);
$\mathcal{M}=1.57\pm0.22$ (NE, r=90~kpc); 
which are consistent with a shock origin.
We note that by estimating the Mach numbers from the density jumps 
as shown in eq. \ref{eq:mach}, we are assuming they are due to shocks. 
However, it could be that these density discontinuities are caused by sloshing fronts, 
in which case the temperature discontinuity would be the reciprocal of the density jump. Deeper X-ray observations would certainly clarify what is the physical origin of the discontinuities.

\begin{figure*}[t]
	\centering	
	
	\includegraphics[width=\textwidth]{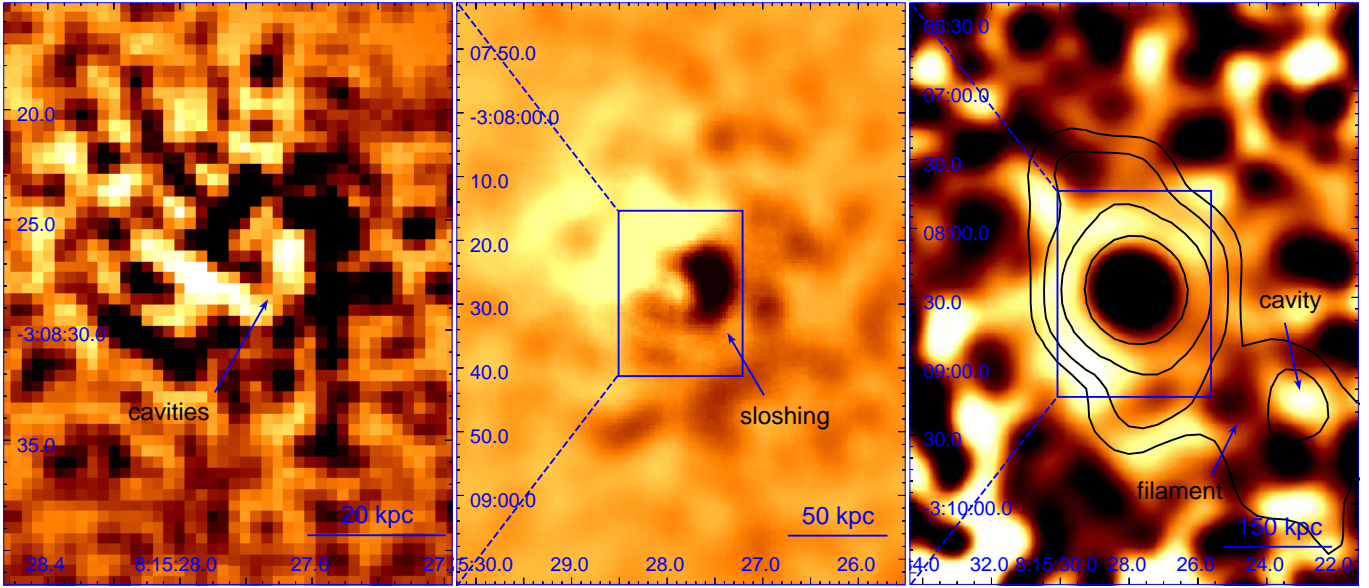}
	\caption{Residuals images of the 0.7-2~keV energy band, with $0\farcs492$ pixel scale. Excess emission appears in black while depressions are shown in white. 
	\textit{Left:} Residual image produced by the subtraction of a $\beta$-model, FoV is 20$\times$25~arcsec$^2$ (65$\times$80~kpc$^2$). The butterfly-shaped cavity is clearly seen in the residuals, together with a dark sharp shell that surrounds it. This dark edge is $\sim$10~kpc long in projection. 
	\textit{Center:} The same image is smoothed with a 10-arcsec Gaussian kernel. FoV is $\sim$1$\times$1.5~arcmin$^2$. Asymmetries in the surface brightness are seen together with a spiral feature, indicative of gas sloshing. 
	\textit{Right:} Unsharp mask image produced by the 
	subtraction of 
	30-arcsec smoothed image from the 15-arcsec smoothed image. The subtracted data are then divided by the sum of the two smoothed images. 
	Field of view is 3$\times$4~arcmin$^2$.
	GMRT 147~MHz contours are overplotted in black (see Fig. \ref{fig:xray} for details). This edge-enhancement method highlights both the outer cavity, co-spatial with the low-frequency radio emission, and the surrounding filament.}
	\label{fig:edges}
\end{figure*}

\subsection{Cavities, Asymmetries and Spiral features}\label{sss:edges} 
Surface brightness edges may be
caused by at least three phenomena: cold fronts,
sloshing of the gas in the central regions of clusters induced by
minor mergers, and shocks either AGN- or merger-driven.
In particular, gas sloshing is signalled by 
the presence of a spiral
structure in the hot gas distribution \citep{markevitch07}. 
To further enhance small amplitude features in the surface brightness, that are otherwise overwhelmed 
by the gradient associated with the cluster, 
we adopt two edge-enhancement methods, presented in Fig. \ref{fig:edges}.
We caution the reader to not over-interpret the structures in Fig. \ref{fig:edges}. Indeed, the two edge-enhancement methods adopted are inherently noisy, and introduce artefacts that complicate quantitative morphological analysis.
	Fig. \ref{fig:edges} (particularly the rightmost panel) is presented mostly as a viewing aid for the major morphological features, which are (almost) already visible from the unprocessed data shown in Fig. \ref{fig:xray}.

To highlight the small-scale ripples in the surface brightness, searching for cavities, edges and 
sloshing features, we produced a residual image   
by subtracting a 2D elliptical $\beta$-model from the 
0.7-2~keV \chn\ image after excising the point sources 
(see Sect. \ref{sss:imaging}), with native pixel scale $0\farcs492$, up to $2\arcmin$.
The left panel of Fig. \ref{fig:edges} shows the residual image, where 
regions of surface brightness excess over the subtracted $\beta$-model
appear in black, whereas deficits are shown in white.
The left panel shows the central 20$\times$25~arcsec$^2$ (65$\times$80~kpc$^2$)
residual image. The butterfly-shaped cavity is clearly seen as a deficit in the residuals. 
The dark sharp region surrounding the cavity to the W, S and SE is seen as an excess. 
This edge has a projected radial extent of $\sim$10~kpc ($\sim$~3$\arcsec$)
with the outer border at a distance of $\sim$20~kpc. This is consistent with the first discontinuity in the slope of the surface brightness radial profiles (see dotted lines at $\sim$20~kpc in Fig. \ref{fig2}).

The central panel shows a larger region, with a FoV of $\sim$1$\times$1.5 arcmin$^2$ ($\sim$215$\times$290~kpc$^2$). Pixel scale is $0\farcs492$, smoothed with a Gaussian kernel of 10$\arcsec$. 
The Gaussian-smoothed residual image shows asymmetries in the surface brightness.
Residuals asymmetric features are expected in case of gas sloshing, and indeed
a spiral feature indicative of sloshing is clearly evident both as excess emission (W and S, dark) and deficit (E and N, white). 
The dark spiral arm is $\sim$6$\arcsec$ wide and $\sim$12$\arcsec$ long, 
which means $\sim$20$\times$40~kpc$^2$. 

We also employed an unsharp mask procedure, which is usually adopted as a bandpass spatial filter, in order to enhance structures between the desired spatial scales. The 0.7-2~keV point-source subtracted image is convolved with Gaussian kernels of different sizes. 
We adopted the kernels of 15$\arcsec$ and 30$\arcsec$ to highlight the 
underlying density discontinuity at $\sim90$~kpc indicated by the change of slope 
	in the surface brightness radial profiles in Fig. \ref{fig2}.
The 30-arcsec smoothed image is subtracted from the 15-arcsec smoothed image
and then the subtracted data are divided by the sum of the two images.
The right panel of Fig. \ref{fig:edges} shows the 
resulting unsharp mask image together with the GMRT 147~MHz contours overlaid in black. The FoV is 3$\times$4~arcmin$^2$.
Asymmetries in the surface brightness are enhanced even at such scales.
The large outer cavity mentioned in Fig. \ref{fig:xray} is clearly seen 
in this unsharp masked image, as indicated by the arrow in Fig. \ref{fig:edges}, as well as the filament that surrounds this cavity.

\section{Results}\label{s:res}	
\subsection{Temperature gradient}
The spectral analysis performed on the \chn\ data outlined in Sect. \ref{sss:spec} 
implies that the cluster is well fitted by an APEC plasma having $\sim$4.2~keV temperature. 
Inverting the $M_{500} -T$ relation of \citet{lovisari15},
this temperature corresponds to an $R_{500}\simeq 0.9$~Mpc, which is consistent with the estimate 
reported by \citet{piffaretti11}.
The temperature analysis 
shows that the
region called \textit{core 1} in Table \ref{tab:xfit} has a cooler temperature of $\sim$ 3.2~keV.
The best-fit gas temperature further drops in the region \textit{core 0}, i.e. considering only the central 10$\arcsec$, to $\sim$ 2.8~keV. 
\footnote{The difference between the two core best-fit temperatures is statistically marginal, $\sim1\sigma$.}
In the four outer sectors (NW, NE, SE, SW), the 
best-fit temperatures are higher (at $\gtrsim2\sigma$ with respect to the region \textit{core 0}).
This result suggests that the cluster could be a cool core cluster, with a lower
temperature in the central region with respect to the peripheral zones.  
In the SW direction (coincident with the low-frequency 147~MHz peak emission) 
there is a hint of a rise in the gas temperature ($\sim 7.5$~keV, see Table \ref{tab:xfit}) with respect to the other three regions. However 
this rise in the SW is not statistically significant ($\lesssim1.5\sigma$),
due to the large uncertainties in the best-fit gas temperatures in the 
cluster outskirts. 
Indeed, the \chn\ image has insufficient 
counts to 
accurately constrain the temperature in the cluster outskirts, 
with current 1-$\sigma$ uncertainties of $\sim$ 20\%.

\subsection{Central density, entropy and cooling time}
Under the assumption
of spherical symmetry 
and that the projected 2D surface brightness can be 
described by a $\beta$-model (see Sect. \ref{sss:sb}),
we can derive the underlying 3D electron gas density distribution
\begin{equation}
n_e (r) = n_{e,0} \left[ 1+ \left(\frac{r}{r_c} \right)^2 \right]^{-\frac{3}{2}\beta} \, .
\end{equation}
The central electron density $n_{e,0}$ is obtained using the
spectral fit. 
When fitting a spectrum with an
$\tt{APEC}$ model, the normalization $K$ gives us the 
electron $n_e$ and proton densities $n_p$ 
\begin{equation}\label{eq:K}
K = \frac{10^{-14}}{4\pi D_A^2 (1+z)^2} \int n_e(r) n_p(r) dV \, ,
\end{equation}
where $D_A$ is the source angular distance.
The normalization in equation \ref{eq:K} is computed 
	in the volume $dV$
	through an infinite cylinder, 
	first integrating radially up to the extraction radius, and 
	then between $-\infty$ and $\infty$
	in order to consider all the emission along the 
	line of sight.
Assuming a fully ionized gas, $n_p\simeq0.82n_e$,
we can recover the gas density. 
The average central density, estimated from the best-fitting parameters of the $\beta$-model describing the
azimuthal profile inside 2$\arcmin$ (see bottom-panel of Fig. \ref{fig2}), is $n_{e,0}\simeq 0.098$~cm$^{-3}$. 

We can now estimate the core entropy index, which is defined as $T/n_e^{2/3}$, 
with the temperature in keV units and the density measured in $cm^{-3}$. 
The entropy can be uniquely determined from the entropy index.
We use the average temperature of the central region \textit{core 0}, i.e. $kT(r<10\arcsec)\simeq 2.8$~keV, and 
the aforementioned estimate of the central density $n_{e,0}$.
The resulting core central entropy index is $\sim$ 13~keV~cm$^2$, and thus the system fits the definition of cool core cluster, taking the characteristic dividing line at $\sim$ 25~keV~cm$^2$ \citep{hudson10}. 
A low central entropy index value is supported by our analysis which 
pinpoints a variety of multiphase gas. 
Indeed, multiphase gas and star formation activity, as hinted for instance by the H$\alpha$ emission (see Fig. \ref{fig:halpha}), 
is expected to be enhanced in BCGs that harbour radio galaxies and with central entropy index below 30~keV~cm$^2$ \citep{cavagnolo08}.
However, note that the position of the AGN is not obvious so its contribution to the X-ray emission cannot be subtracted resulting in an overestimate of the X-ray emission, and consequently the central density and entropy.
The flux from the cool core contained within $0.15R_{500}$ is
	 $\sim$1.5$\times$10$^{-12}$~erg~cm$^{-2}$~s$^{-1}$ 
in the 0.1-2.4~keV band, corresponding to a luminosity 
of $\sim$1.5$\times$10$^{44}$~erg~s$^{-1}$.
The upper limit on the X-ray nuclear flux (observed-frame and uncorrected for absorption), computed considering 
all the emission coming from the central circle with radius 2$\arcsec$, is $\sim$3$\times$10$^{-15}$~erg~cm$^{-2}$~s$^{-1}$ in the soft 0.5-1~keV and medium 1-2~keV bands, and $\sim$7$\times$10$^{-15}$~erg~cm$^{-2}$~s$^{-1}$
in the hard 2-7~keV band. 
Nonetheless, the galaxy is classified in the optical as a LERG \citep{buttiglione10}, and as such the X-ray emission from the accretion disk is 
	expected to be low \citep[e.g.,][]{hardcastle09}, hence the AGN X-ray contribution probably does not affect much our results.

Additional evidence for a cool core cluster is given by the cooling time
\begin{equation}
t_{cool} = \frac{3}{2} \frac{(n_e + n_p) \, kT}{\Lambda(T,Z) \, n_e n_p} \,
\end{equation}
where $\Lambda(T,Z)$ is the so-called cooling function, which can be estimated 
using $\tt{XSPEC}$. 
We found that $\Lambda(kT =2.8$~keV$, Z=0.3) \simeq 1 \times 10^{-23}$~erg~cm$^3$~s$^{-1}$, 
which gives a cooling time $t_{cool} \sim 5 \times 10^{8}$~yr 
(assuming 
an APEC model with best-fit parameters obtained for the region \textit{core 0}, with 0.3 solar abundance, and a fully ionized gas).
The age of the Universe at redshift 0.198 is $t_{age}=11.122 \times 10^{9}$~yr, thus $t_{cool}/t_{age}<0.05$ supports the 
cool core nature of the cluster.\\

\subsection{Cavities energetics and age}\label{s:disc}

Fig. \ref{fig:xray} shows the 8~ks \chn\ snapshot observation and the 8.4~GHz radio contours of the core region. 
There is a wealth of structures
already in the snapshot observation, which suggests a complex dynamical state of the system, 
that could have 
experienced multiple AGN outbursts. 
The luminosity $L_{500}$ extrapolated up to $R_{500}$ in the 0.1-2.4~keV band is $\sim$2.8 $\times$~10$^{44}$~erg~s$^{-1}$, which is consistent with the \textit{ROSAT} measurement \citep{piffaretti11}. 
As already introduced in Sect. \ref{sss:imaging}, in the inner region 
there is a ``butterfly-shaped'' cavity wrapping around the SW jet/lobe.
The steep (radio) spectrum ``S-lobe'' impinges on part of this cavity.  
This system of cavities can be approximated as two contiguous ellipsoids,
at a distance $\sim$~10~kpc ($\simeq3\arcsec$), 
each with a semi-major axis of 2$\arcsec$ and a semi-minor axis of 1$\arcsec$. 
To compute the volume we will assume that the semi-axis along the line of sight is the 
average of the two, i.e. 1$\farcs$5. 
The resulting volume of each ellipsoid is $1.2\times10^{67}$~cm$^3$, corresponding to 
a total cavity volume of $\sim820$~kpc$^3$.

On large scales, the low-frequency 147 MHz emission extends
for hundreds of kpc 
to the SW.
As already discussed in Sect. \ref{sss:imaging}, this low-frequency 
radio emission is aligned with the inner scale jet emission. 
As highlighted in Fig. \ref{fig:xray} there is a large X-ray filament extending towards the south-west, which 
spans $\sim$2$\arcmin$ ($\sim$~384~kpc, projected). It could also be the southern border of a large cavity
in the hot gas, appearing as a deficit in the
X-ray surface brightness, 
which is co-spatial with the peak of 147 MHz radio emission (see  
Fig. \ref{fig:xray}, right panel). 
This cavity is marked by a blue dashed ellipsoid with semi-major axis 14$\arcsec$ and semi-minor axis 13$\arcsec$ 
(the semi axis along the line of sight is taken 13$\farcs$5). 
The centre of this ellipsoid lies at a (projected) distance of $\sim$90$\arcsec$ from the AGN. 
The volume of this cavity would be $\sim10^{70}$~cm$^3$, corresponding roughly to 
$3\times10^5$~kpc$^3$. 
In both cases, the estimated cavity volumes do not take into account any effect of projection, therefore they should 
be considered as upper limits on the real cavity volume. Also, the choice of the semi axis along the line of 
sight is arbitrary, even though it is consistent with the standard approximation usually 
adopted in the literature.

The pressure associated with each cavity is given by $P_{cav}=(n_e + n_p) kT \simeq 1.82 n_e(r) kT(r)$.
By assuming
that: i) the electron density is described by the $\beta$-model parameters derived from the 
azimuthal fit of the region contained in 2$\arcmin$ (see Fig. \ref{fig2});
ii) a fully ionised gas; and iii) taking $kT$=2.8~keV as determined by the spectral fitting in the central 10$\arcsec$ region (see Table \ref{tab:xfit}) for the inner cavity and instead $kT$=4.2~keV for the cavity that is located at $\sim$90$\arcsec$;
the resulting pressures are $P_{cav}^{in}\simeq 0.5 $~keV~cm$^{-3}$ 
for the butterfly-shaped cavity and 
$P_{cav}^{out} \simeq 0.05$~keV~cm$^{-3}$ 
for the outer cavity.

Assuming a relativistic plasma, 
the minimum energy required to create a cavity, e.g. the 
	cavity enthalpy, is 
\begin{equation}\label{eq:enthalpy}
E_{cav} =\frac{\gamma}{\gamma -1}\, P_{cav} V = 4 P_{cav} V
\end{equation}
which is $\sim7 \times 10^{58}$~erg 
for the inner cavity and $\sim 3 \times 10^{60}$~erg 
for the outer cavity.

If the low frequency radio emission comes from a buoyantly rising radio 
	lobe that originated in a past AGN outburst and has displaced the X-ray emitting gas, 
we can compute its age given its projected cluster-centric 
distance
$R\sim 90\arcsec$ ($\sim290$~kpc). 
Assuming the bubble is launched from the nucleus and rises in the plane of the sky at the sound speed $c_s$ \citep[see e.g. the review from][]{mcnamaranulsen07}
\begin{equation}
c_s = \sqrt{\frac{kT}{\mu m_p}} \simeq 1100 \left(\frac{kT}{5 \, \rm{keV}}\right)^{1/2} \,\, \rm{km~s}^{-1}
\end{equation}
where $\mu= 0.62$ is the mean molecular weight in units of the proton mass $m_p$, and that the ambient temperature at distance $R$ is $kT(R) = 4.2$~keV,
then the age of this radio emitting plasma would be 
\begin{equation}\label{eq:tc}
t_{rise} \simeq R/c_s = 280 \, \rm{Myr} \, .
\end{equation}
The simple assumption of a sonic expansion of the bubble does not 
consider an evolution in the bubble velocity, 
where the initial stages of the cavity inflation are usually thought to be 
supersonic and then followed by subsonic buoyant rise. 
Nonetheless, eq. \ref{eq:tc} represents a simple approach that can 
be used to get an estimate of age
average between the two inflation phases.\\
The age of this radio emission should be of the same order as the AGN radio phase, if this radio emission is related to previous radio outbursts.
A timescale of $\sim$~10$^8$~yr is indeed consistent with cycling times between the triggering of radio activity, the onset of quiescence and the subsequent re-ignition of activity, e.g. 10$^7$-10$^8$~yr 
\citep{parma99, best05, shabala08, tremblay10}. 
In general, synchrotron losses limit radio source lifetimes to $\sim$10$^8$~yr in a few $\mu$G magnetic field typical of lobes in radio galaxies, unless there has been re-acceleration of the electron population.

We can estimate the same quantities also for the inner 
	cavity, at a distance of $\sim$~10~kpc, assuming that the temperature inside the cavity is given by the 
	average temperature $kT\sim2.8$~keV we found in the region \textit{core 0}. 
	We then get an age of $\sim12$~Myrs. 

We can finally compute the jet power 
$P_{jet}$, which is the minimum work required to inflate
a cavity with a volume $V$, i.e. the cavity enthalpy in eq. \ref{eq:enthalpy}, divided by the age of the
cavity, 
\begin{equation}
P_{jet} = 4 PV/t = E_{cav} / t~.
\end{equation}
As shown in eq. \ref{eq:tc}, the cavity age can be approximated by the bubble rise time, 
under the (simplistic) assumption of 
sonic expansion. 
The resulting jet powers are $\sim1.9\times10^{44}$~erg~s$^{-1}$ for the inner cavity and
$\sim3.4\times10^{44}$~erg~s$^{-1}$, which are consistently measured also in other galaxy clusters 
\citep[see the review of ][their Fig. 7]{bykov15}.

\section{Summary and Conclusions}\label{s:concl}

Performing a multi-wavelength analysis from publicly available radio, optical and X-ray images, we have investigated the nature of the multiphase emitting gas both in the nuclear and surrounding regions of 3C 196.1.
The emerging scenario is that this radio galaxy could have undergone several radio outbursts on multiple epochs and its cluster may have experienced a merger. In particular: 
\begin{enumerate}
		\item Optical HST observations reveal clear signs of interactions between the radio jet and lobe with the 
		\ha\ emitting gas. 
		Distortions in the optical isophotes of the BCG support a scenario of merging. 
		In Fig. \ref{fig:halpha} there is filamentary \ha\ emission extending eastwards, draped across the radio lobe. 
		This \ha\ filament must be dynamically very short lived. 
		Indeed, recombination times for	T~$\sim$~10$^4$~K gas are of order 10$^3$~yr \citep{osterbrock06}, while lifetimes of radio sources are of order 10$^8$~yr. The process by which the \ha\ gas is ionized should therefore be ongoing throughout the lifetime of the radio source. This implies a strong connection between AGN activity and observed emission line properties;
					
	\item The low-frequency and high-frequency radio data helped us to investigate both the inner and outer scale of the cluster. By comparing the galactic-scale radio and HST images, we have found 
	that the \ha\ emission is aligned with the inner scale jet, an example of the well-known ``alignment effect'' \citep{fosbury86,hansen87,baum88,baum90,devries99,tremblay09}. 
	The combination of the radio and the X-ray images allowed us to discover 
	cavities located in the galaxy, 
	at $\sim$~10~kpc, and in the cluster outskirts, at $\sim$~290~kpc, whose presence suggests past AGN outbursts;

	\item The \ha\ emission is bounded to the SW by the butterfly-shaped cavity in the inner region and the filamentary emission to the NE is co-spatial with the northern hotspot;

	\item The \chn\ X-ray data analysis allowed us to constrain basic physical parameters of the cluster and of the ICM gas. 
	We revealed the presence of discontinuities in the gas density by analysing the 
	surface brightness profiles. 
	Assuming these discontinuities are due to shocks, we determined the Mach numbers 
	from the density jumps, derived under the Rankine-Hugoniot conditions. 
	The inferred Mach numbers and density jumps are consistent with being originated by shocks. 
	We also employed two different edge-enhancement methods 
	to better visualize the small ripples in surface brightness associated with density discontinuities. 
	In this way we found a spiral pattern, characteristic of gas sloshing, suggesting a past merger event;
	
	\item We found a decrease in temperature in the inner region, suggesting that 
	3C~196.1 is hosted in a 
	cool core cluster. This result is further supported by the analysis of the X-ray surface brightness, through which we determined the core entropy index of $\sim$~13~keV~cm$^2$ and the cooling time $t_{cool} \sim 500$~Myrs;

	\item Finally we computed the volumes, pressures and enthalpies $E_{cav}$ associated with the cavities: $E_{cav}\sim7 \times 10^{58}$~erg for the inner cavity and $E_{cav}\sim 3 \times 10^{60}$~erg
	for the outer cavity. 
	We also determined the 
	lifetimes of both cavities, the inner cavity of $\sim12$~Myrs and the outer one of
	$\sim$280~Myrs, by assuming that the radio plasma is 
	a sonic rising bubble. This lifetime is consistent with radio cycle lifetimes and thus supports 
	the origin of the outer cavity as a past AGN outburst. 
	With the estimate of bubble rise time, we computed the jet powers $P_{jet}$ and found 
		$P_{jet}\sim1.9\times10^{44}$~erg~s$^{-1}$ for the inner cavity and $P_{jet}\sim3.4\times10^{44}$~erg~s$^{-1}$ for the outer cavity.
	
\end{enumerate}

Given that the 
\ha\ 
gas filaments are aligned with the radio jets/bubbles, we are probably witnessing an example here of lobes uplifting cold gas, 
even though robust conclusions cannot be drawn since we do not have any kinematic information. 
Deeper X-ray and radio data would be critical to provide clues about the dynamics of the cold and hot phase gas,
and to obtain a more complete understanding of the morphology and of the nature of the system.
Additional deeper data would be also needed to 
better define cavities, constrain the temperature and metallicity distributions,
and look for additional merger features.
This source thus represents an intriguing opportunity 
to quantify, with deeper X-ray observations, the impact of AGN activity in the 
outskirts of a group in a   
nearby cluster.

\acknowledgements
	We thank the referee for valuable suggestions that improved the quality of the manuscript.
	We acknowledge C. C. Cheung for kindly providing the 8.4 GHz radio map. 
	This investigation is supported by the National Aeronautics and Space Administration (NASA) grants GO4-15096X, GO4-15097X and GO6-17081X. 
	FR acknowledges support from FONDECYT Postdoctorado 3180506 and CONICYT Chile grant Basal-CATA PFB-06/2007.
	LL acknowledges support from NASA through contract NNX17AD83G.
	This work is supported by the "Departments of Excellence 2018 - 2022" Grant awarded by the Italian Ministry of Education, University and Research (MIUR) (L. 232/2016). This research has made use of resources provided by the Compagnia di San Paolo for the grant awarded on the BLENV project (S1618\_L1\_MASF\_01) and by the Ministry of Education, Universities and Research for the grant MASF\_FFABR\_17\_01. FM acknowledges financial contribution from the agreement ASI-INAF n.2017-14-H.0.
	The work of SB and CO was supported by NSERC (Natural Sciences and Engineering Research Council of Canada).
	This research has made use of NASA’s Astrophysics Data System; SAOImage DS9, developed by the Smithsonian Astrophysical Observatory, and the NASA/IPAC Extragalactic Database (NED), which is operated by the Jet
	Propulsion Laboratory, California Institute of Technology, under contract with the National Aeronautics and Space Administration. 
	The (USA) National Radio Astronomy Observatory (NRAO) is operated by Associated Universities, Inc. and is a Facility of the (USA) National Science Foundation.
	We thank the staff of the GMRT that made these observations possible. GMRT is run by the National Centre for Radio Astrophysics of the Tata Institute of Fundamental Research.
	Facilities: VLA, GMRT, HST, CXO (ACIS).

%
\bibliographystyle{aasjournal} 
\bibliography{ref} 
%

\end{document}